\documentclass[reprint,amsmath,amssymb,aps,twocolumn,superscriptaddress]{revtex4-2}

\usepackage{graphicx}
\usepackage{dcolumn}
\usepackage{bm}
\usepackage[caption=false]{subfig}
\usepackage{float}
\usepackage{placeins}
\usepackage{xcolor}
\usepackage{siunitx}
\usepackage{titlesec}
\usepackage{braket}
\usepackage[colorlinks=True,linkcolor=red,citecolor=blue,urlcolor=blue]{hyperref}

\usepackage[ngerman,english]{babel}
\defineshorthand{"~}{\babelhyphen{nobreak}} 
\useshorthands{"}

\begin{document}
\preprint{V0.0}

\title{Overlapping Top Gate Electrodes based on Low Temperature Atomic Layer Deposition for Nanoscale Ambipolar Lateral Junctions
}

\author{C. Fuchs$^{1,2}$, L. Fürst$^{1,2}$, H. Buhmann$^{1,2}$, J. Kleinlein$^{1,2,\ast}$, and L. W. Molenkamp$^{1,2}$\\
\normalsize{$^{1}$Physikalisches Institut (EP3), Universit\"{a}t W\"{u}rzburg, Am Hubland, 97074 W\"{u}rzburg, Germany}\\
\normalsize{$^{2}$Institute for Topological Insulators, Am Hubland, 97074 W\"{u}rzburg, Germany}\\
\normalsize{$^\ast$To whom correspondence should be addressed; E-mail: johannes.kleinlein@physik.uni-wuerzburg.de}
}
\date{\today}

\begin{abstract}
We present \textit{overlapping top gate electrodes} for the formation of gate defined lateral junctions in semiconducting layers as an alternative to the back gate/top gate combination and to the split gate configuration. The optical lithography microfabrication of the overlapping top gates is based on multiple layers of low-temperature atomic layer deposited hafnium oxide, which acts as a gate dielectric and as a robust insulating layer between two overlapping gate electrodes exhibiting a large dielectric breakdown field of $>1 \cdot 10^9\,\text{V}/\text{m}$. The advantage of overlapping gates over the split gate approach is confirmed in model calculations of the electrostatics of the gate stack. The overlapping gate process is applied to Hall bar devices of mercury telluride in order to study the interaction of different quantum Hall states in the $nn'$, $np$, $pn$ and $pp'$ regime.
\end{abstract}

\maketitle

\section{Introduction}

With the advent of two"~dimensional transport materials, purely gate defined ambipolar lateral junctions of low-density electronic systems have attracted increasing attention because of their interesting physical properties and potential applications in {(opto"~)}electronics (see for example {Refs.}\,\cite{Williams.2007,Wang.2021, Baugher.2014, Ross.2014, Pospischil.2014}).
In these works, two zones of differing charge carrier density and type are induced  either by a combination of a \textit{top} and a \textit{back gate} (e.g. Ref.\,\cite{Williams.2007}) or a \textit{split gate} \cite{Thornton.1986, Baugher.2014}.
While a top gate is deposited on top of the semiconductor stack, a back gate consists usually of a conducting substrate or a conducting film within the layer stack of the semiconductor.
Whereas such back gates act on the entire structure, the top gates are most flexible for local gating, as they can be shaped lithographically.
The closest lateral spacing of two neighboring top gate electrodes to form a split gate is typically in the range of about 100\,nm (e.g. Ref.\,\cite{Baugher.2014}).
The interplay of quantum Hall edge states along a lateral junction is of special interest and has been investigated thoroughly in gallium arsenide (see Ref.\,\cite{Haug.1993} for a review), and more recently in graphene \cite{Williams.2007, Abanin.2007,Amet.2014}.
In graphene, $n$"~ and $p$"~type co"~propagating quantum Hall edge channels can be studied, a regime inaccessible for gallium arsenide.

Gate defined lateral junctions in the ambipolar two-dimensional topological insulator mercury telluride \cite{Konig.2007, Roth.2009} have been used to probe the interaction of quantum Hall (chiral) and quantum spin Hall (helical) edge channels \cite{Gusev.2013, Calvo.2017}.
\citet{Gusev.2013} combine a $n$ doped quantum well with a top gated strip, which forms a $nxn$ junction ($x = n', i, p$).
Thereby, only the density of the strip is tunable, resulting in a partially gate defined junction.
\citet{Calvo.2017} employ a combination of a global back gate and a top gated strip, forming a fully gate tunable junction ($nn'n$, $npn$, $pp'p$ and $pnp$).
While the device generally works in all four regimes, the experimentally observed quantum Hall transport through the lateral junction is distorted in all quadrants when compared to the theoretical expectation value.
A likely reason for this behavior is the increased disorder of the mercury telluride quantum wells when a gallium arsenide substrate is used as a back gate in the layer stack \cite{Schlereth.2020}, caused by the large lattice constant mismatch between the substrate and the mercury cadmium telluride epitaxial layers, which makes coherent studies of homogeneous junctions much more challenging.
In addition, both works employ ion beam etching, which is known to strongly reduce device quality \cite{Bendias.2018}.

In this work, we present an approach to use multiple lithographically defined top gates to minimize the ungated area between neighboring split gate electrodes by overlapping them, a design we call \textit{overlapping top gates}.
A similar  top gate design has previously been applied to contact low electron density field effect transistors \cite{Beveren.2010} and more recently for reconfigurable devices \cite{Pang.2019}.
We  expand the idea to  a low-temperature overlapping gates process based on atomic layer deposition of hafnium oxide, that is suitable for the heat sensitive mercury telluride quantum wells (Sec.\,\ref{design}). 
Devices fabricated using the process are characterized for dielectric breakdown and gate action.
Sec.\,\ref{sim} quantifies the advantage of the overlapping gate design over the conventional split gate approach  by calculating the electric field under and between the gates, along with the width of the gate defined lateral junction.
In Sec.\,\ref{QHE}, we introduce a device that features a lateral junction based on overlapping gates for the investigation of the quantum Hall effect when the junction is in bi"~ or unipolar configuration (four quadrants $nn'n$, $npn$, $pp'p$ and $pnp$). 
The overlapping gates allows to avoid a back gate, so higher"~quality quantum wells on an insulating substrate can be employed.
As a result, an excellent Hall quantization is observed, exhibiting quantization in all four quadrants.

\begin{figure*}[t]
    \centering
    \includegraphics[]{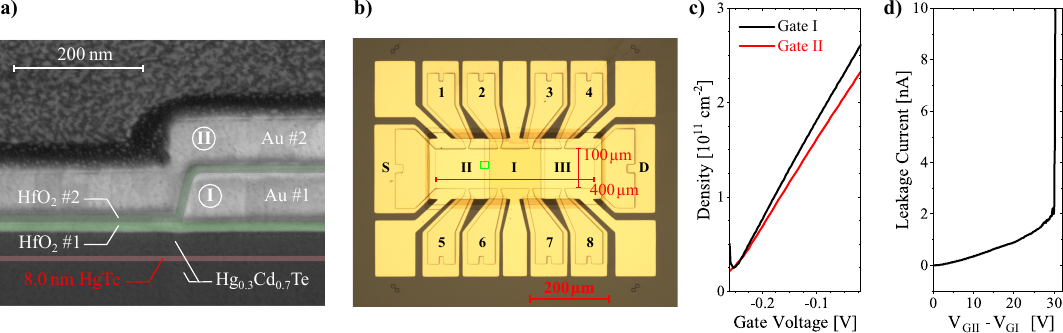} 
    \caption{(a) A scanning electron microscope cross-sectional view of the overlapping gate structure. The two 14.5\,nm thick hafnium oxide dielectric layers are colored in different shades of green. The dark shadow under the gold electrodes is the Ti adhesion layer.  
    (b) A micrograph of the 10-terminal Hall bar device that employs the overlapping gate. The major dimensions, contact numbers and region/gate labels are indicated. The green box indicates where (a) is located in the device.
    (c) The neighboring gate I and gate II reveal a linear dependence of carrier density to gate voltage, indicative of a constant gate action. (d) The leakage current through the dielectric layer when a gate voltage difference between neighboring gates is applied (measured in a test device as described in the main text). 
    }
    \label{Fig:1}
\end{figure*}

\section{Overlapping Top Gate Electrodes}
\label{design}

The overall idea of the overlapping top gate electrodes is the following: a device is covered by a thin film of dielectric, deposited by atomic layer deposition. 
After the deposition of this insulating layer, the first gate electrode is lithographically defined and deposited.
Then, a second layer of dielectric is deposited over the device.
Since atomic layer deposition is a conformal coating process, the side faces of the first gate are uniformly covered.
Subsequently, a second gate electrode is defined by (optical or electron-beam) lithography and metallized with a small lateral overlap of a few micrometer with the underlying gate electrode.
As a result, the gate electrode separation in the overlapping region is purely defined by the thickness of the dielectric, typically $\sim 14.5$\,nm \cite{Shekhar.2022}, which is well below the smallest separation that can be reliably achieved for lithographically defined split gates, as will be discussed in detail later.
Generally, the overlapping gate approach is compatible to any dielectrics that can be deposited by atomic layer deposition.

We apply the overlapping gate design on a molecular beam epitaxy grown $\text{Hg}_{0.3}\text{Cd}_{0.7}\text{Te}$/HgTe/$\text{Hg}_{0.3}\text{Cd}_{0.7}\text{Te}$ quantum well structure.
The underlying substrate is a commercially available $\text{Cd}_{0.96}\text{Zn}_{0.04}\text{Te}$ wafer, which is undoped and insulating.
The latter also applies to the top and bottom mercury cadmium telluride barrier layers, that sandwich the 8\,nm thick mercury telluride quantum well.
Fig.\,\ref{Fig:1}a shows a cross sectional image of the overlapping gate structure on top of the mercury telluride quantum well.
Here,  hafnium oxide grown by a low"~temperature atomic layer deposition process is employed as a dielectric.
The thickness of both hafnium oxide layers is $\sim$\,\SI{14.5}{nm} (90 growth cycles), and each gate electrode consists of a titanium adhesion layer ($\sim$\,\SI{2}{nm}) and $\sim$\,\SI{70}{nm} gold.
Before the deposition of the dielectric or the metal layers -- {i.e.} after each lithography step -- the surface of the device is cleaned for 3\,min in a gentle oxygen plasma.
The process results in clearly separated gate electrodes with a uniform separation of $\sim$\,\SI{14.5}{nm}.

To demonstrate the functionality of the overlapping top gates, we built a 10-terminal Hall bar device, shown in Fig.\,\ref{Fig:1}b.
The fabrication is  based entirely  on optical lithography.
First, the mesa is defined by wet-chemical etching \cite{Bendias.2018}.
Ten ohmic contacts (source S, drain D, and voltage probes 1 to 8) are fabricated by removal of the top $\text{Hg}_{0.3}\text{Cd}_{0.7}\text{Te}$ barrier using argon ion milling and \textit{in situ} deposition of 50\,nm of eutectic gold"~germanium, covered with 50\,nm gold.
Subsequently, the overlapping gate stack is deposited to fabricate three gates: After the first hafnium oxide layer the lower gate electrode I is deposited, and after the second layer of hafnium oxide the upper gates II and III.
The gates overlap the areas of the  ohmic contacts by a few micrometers to ensure a homogeneous carrier density within the entire sample area.
Note that the two hafnium oxide layers can be identified as slightly darkened rectangles around the Hall bar in Fig.\,\ref{Fig:1}b.
The device is electrically contacted by wire"~bonding.

The carrier density under gate I and  gate II (the mirror"~symmetric gate III acts similarly as gate II) are determined by the low field Hall resistance, which is measured applying low frequency lock"~in techniques at liquid helium temperature.
The dependence of the charge carrier density $n$ on gate voltage $V_G$ in the  gated regions is shown in Fig.\,\ref{Fig:1}c.
Both gates allow for a linear tuning of the electron density  to full depletion.
As gate I employs a thinner dielectric layer than gate II, the gate action $\frac{\Delta n}{\Delta V_G}$ is about 10\% larger.
The carrier density can be tuned independently under both gates, and changing the density under one of them does not effect the density under the other (no crosstalk).

To test the breakdown voltage of the dielectric layer between the gates, a specially designed test device with seven overlapping gates (not shown) is built.
We sweep the voltage between two neighboring gate electrodes and measure the leakage current using a source-measure unit with pA resolution (\textit{Keysight B2962B}).   
Fig.\,\ref{Fig:1}d shows a representative I-V-characteristic of such a measurement.
At small voltages the differential resistance is larger than \SI{50}{G\ohm}, and it remains above \SI{10}{G\ohm} until the dielectric breaks down at 30\,V.
All seven gates show a breakdown voltage $>$\SI{15}{V}, with the largest breakdown voltage reaching $38$\,V.
The breakdown electric field for the overlapping gates separated by $\sim 14.5$\,nm hafnium oxide is therefore $> 1 \cdot 10^9\,\text{V}/\text{m}$.
This value is comparable and partly surpasses the breakdown field of well established oxides commonly used for device applications \cite{Verweij.1996}.
In particular, the breakdown field with hafnium oxide as an insulator between the overlapping gates appears larger than with aluminum oxide \cite{Lawrie.2020}.
The large breakdown field is remarkably high given the low temperature (30$\,^\circ$C) at which the hafnium oxide layers is deposited.

\section{Overlapping Gates vs. Split Gates}
\label{sim}

\begin{figure*}[t]
    \centering
    \includegraphics[]{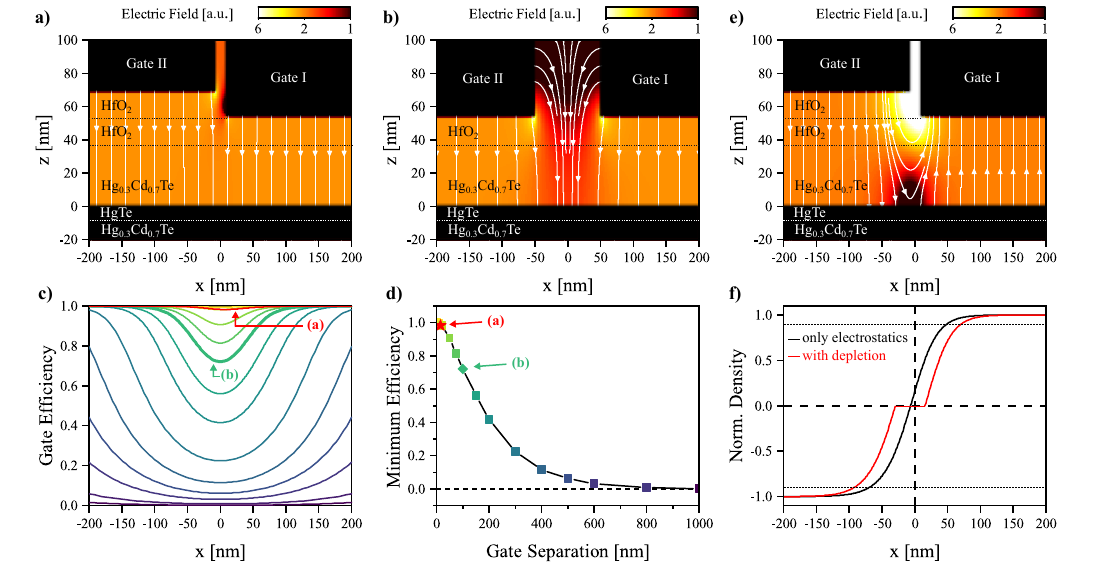} 
    \caption{
    Calculation of electrostatic landscape for (a) overlapping gates and (b) split gates with a separation of 100\,nm, when the electron density far away from the overlap/split is set  to the same value. The color-scale indicates the electric field and the vectors the electric field lines. (c) The gating efficiency in dependence of position relative to the gate overlap/split. The red curve corresponds to the overlap (a) and the thick green curve to the split gate with 100\,nm separation (b). The other curves correspond to various split gate separations between 10\,nm and \SI{1}{\micro m}, which is color mapped to the separations indicated in (d). (d) The minima of the curves in (c) against their respective gate separation. (e) The electrostatic landscape for the overlapping gates when the electron density far away from the overlap is set to the same absolute value with opposite signs, forming a lateral $pn$ junction. (f) The black line indicates the normalized carrier density in the quantum well when (e) applies. The dotted line indicate $\pm 90\%$ of the carrier density far away from the junction.  The red curve includes an estimation of the depletion width of {approx.} 45\,nm.
    }
    \label{Fig:2}
\end{figure*}

A calculation of the electrostatic landscape in the vicinity of the lateral gate defined junction is important to quantify the advantages of the overlapping top gate design to the more conventional split gate approach. 
Below, we calculate the electrostatic potential in the dielectric layer close to the gate defined junction (which determines the gate action next to it) and compare it with a similar calculation for a split gate geometry.

The potential is calculated by solving the Poisson equation with fixed gate electrode and quantum well potentials as an input parameter.
The potential of the gate electrode is defined by the voltage which is applied relative to the quantum well.
The quantum well is modeled by a uniform potential ($=0$\,V in the following).
As the electrostatic landscape is invariant normal to the plane of the cross section depicted in Fig.\,\ref{Fig:1}a, the problem reduces to a 2D Poisson equation.
The latter is solved following Ref.\,\cite{Zaman.2022}, assuming that the dielectric constants of hafnium oxide \cite{Shekhar.2022} and $\text{Hg}_{0.3}\text{Cd}_{0.7}\text{Te}$ \cite{Baars.1972} are the same ($\epsilon_r = 14\pm3$) within the error margin.
The electric field $E$ is the derivative of the potential, and the electric field component $E_z$ normal to the surface of the HgTe layer defines the  electronic carrier density in the quantum well  following $n(x) = \frac{\epsilon_0 \epsilon_r}{e} E_z(x, z = 0^+)$, where $\epsilon_0$ is the vacuum permittivity and $e$ is the electron charge.
Fig.\,\ref{Fig:2}a and Fig.\,\ref{Fig:2}b show the calculated electrostatic landscape in the vicinity of the two gates I and II for the overlapping gate and a split gate with a separation of 100\,nm, respectively.
In both cases, the gate voltages are chosen to result in similar carrier densities under both gates far away from the (unipolar) junction .
The color scale represents the absolute value of the electric field, and the arrows indicate the electric field lines.
The overlapping gate design reveals a much more homogeneous field, also in the overlap area, as compared to the split gate design.
This is quantified in Fig.\,\ref{Fig:2}c, where the gate efficiency, defined as the ratio of the induced carrier density at the lateral position $x$ and the carrier density far away from the junction, is shown for the overlapping gate and split gates with different separations.
The red curve, indicated by (a), is calculated for the overlapping gate geometry, and reveals a very uniform gate efficiency that barely deviates from 1.
In Fig.\,\ref{Fig:2}d, we plot the minimum in gate efficiency for all the curves in Fig.\,\ref{Fig:2}c versus the lateral separation of the neighboring gates.
The gate efficiency of the split gates decreases from 100\% as the separation of the gate electrodes increases, falling to 78\% for 100\,nm gate separation.
When the gate split is increased further, the minimum gate efficiency quickly drops towards zero, until it is $\ll1\%$ for \SI{1}{\micro m}.
In this limit, barely any gate action is observed in the 400\,nm wide window shown in Fig.\,\ref{Fig:2}c.
In comparison, the minimum efficiency of the overlapping gate is 98.2\% (red star), similar to that of  the split gate with a gate separation of $< 20$\,nm.

From a microfabrication perspective, a split gate with a gate separation of 100\,nm is already very challenging to fabricate using electron beam lithography, especially when large aspect ratios of the gate split are required (as is the case in Fig.\,\ref{Fig:1}b where the Hall bar is \SI{100}{\micro m} wide).
Employing optical lithography on mercury telluride quantum wells, a gate separation of \SI{2}{\micro m} can be reliably achieved for the stated structure width, resulting in a practically ungated region between the gates of about \SI{2}{\micro m}.
Compared to that, the overlapping gate process neither requires maxing out the resolution of the optical process, nor does it result in a relevant ungated area.
It rather provides a gate separation of only $\sim 14.5$\,nm (or even smaller when a thinner dielectric is employed),  massively undercutting the limits of electron beam lithography.

A similar calculation provides insights into the electrostatics when the densities under the two neighboring gates differ.
The calculation for a (bipolar) $pn$ junction  is shown in Fig.\,\ref{Fig:2}e, where the absolute value of the charge carrier density far away from the overlap is the same, but with opposite signs (gate II: $p$ type, gate I: $n$ type).
Due to the opposite charge carrier type, the electric field vectors are oriented in opposite directions under the two gates, and approach zero at the quantum well in the vicinity of the overlap.
The corresponding electrostatically induced carrier density distribution is presented by the black curve in Fig.\,\ref{Fig:2}f.
The graph shows a continuous transition in carrier density between the constant values far away from the overlapping gate.
A slight shift of approx. 8\,nm of the otherwise symmetric transition away from gate I can be observed, which is expected due to the increased distance of gate II to the quantum well.
The electrostatic transition width $w_T$ -- here defined by reaching 90\% of the carrier densities far away from the junction -- is evaluated at around 120\,nm.
The electrostatic transition width can be used as an upper bound of the depletion width $w_d$, which defines the size of the region in the $pn$ junction where no free carriers are present.
A more rigorous approximation of the latter includes the band structure of the underlying material.
\citet{Chaves.2023} present an analytical solution for a split gate geometry with a very narrow spacing between the split gates.
In good approximation, their formula for the depletion width 
\begin{equation}
w_d = \dfrac{\pi d}{\text{sech}^{-1}\left( \frac{\pi}{4} \frac{E_G}{2 e V_G}\right)}
\label{Eq1}
\end{equation}
will also model our more complicated overlapping gate geometry, as the electrostatics do not significantly differ to a narrow split gate ($< 20$\,nm, see Fig.\,\ref{Fig:2}c).
In Eq.\,(\ref{Eq1}), $d = 53 - 68\,\text{nm}$ is the thickness of the dielectric layer and $E_G \approx 15\,\text{meV}$ is the quantum well band gap \cite{Novik.2005}.
The model predicts a depletion width of $w_d = (45 \pm 15) \,\text{nm}$ at a gate voltage of $\pm 0.2$\,V (relative to the charge neutrality point, equivalent to a carrier density of $\sim \pm 2 \cdot 10^{11}\,\text{cm}^{-2}$).
The error represents an upper limit estimation.
The depletion width relative to the electrostatic effects is shown in the red curve of Fig.\,\ref{Fig:2}f, where the electrostatic solution is split around zero carrier density by the depletion width.
As the curve does not include self-consistent corrections to the potential landscape when a part of the quantum well is fully depleted, the red curve should only be interpreted as a plausible approximation.
The electrostatic calculation as well as the analytical approximation imply that the gate defined $pn$ junction in our material certainly occurs within less than 120\,nm, significantly undercutting a design based on  split gates defined by optical lithography.

\section{Selective Quantum Hall Edge Channel Transmission}
\label{QHE}

\begin{figure*}[t]
    \centering
    \includegraphics[width = \textwidth]{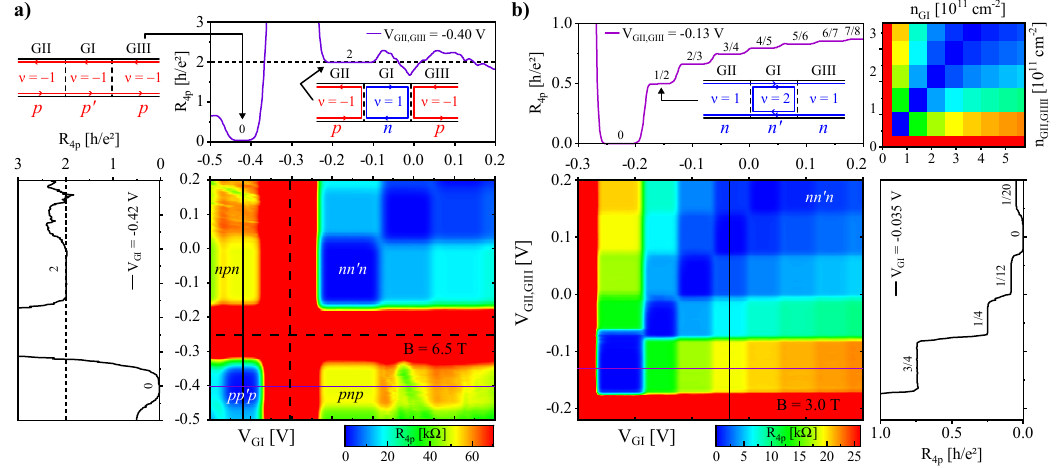} 
    \caption{The resistance $R_{4p}$ in the quantum Hall transport regime depending on gate voltages $V_{GI}$ and $V_{GII\&III}$ applied to gate I and gate II\&III, respectively. (a) The resistance at 6.5\,T, where the junctions are in $nn'n$ (first quadrant), $npn$ (second quadrant), $pp'p$ (third quadrant) and $pnp$ (fourth quadrant) configuration. The line cuts correspond to the accordingly colored  horizontal and vertical solid lines. The insets sketch the edge channel configuration for the indicated regimes. (b) The $nn'n$ quadrant at 3\,T, revealing forty quantized plateaus. A calculation of the quantized resistances over the covered density range is presented the upper right corner.
    }
    \label{Fig:3}
\end{figure*}

In this section, we study magnetotransport through the lateral junctions of the 10-terminal Hall bar (see Fig.\,\ref{Fig:1}b) in the quantum Hall regime.
Generally, a differing number of chiral edge channels on the two sides of a unipolar lateral junction results in selective edge channel transmission, as first observed by \citet{Haug.1988}.
Later, \citet{Williams.2007} have shown that edge channels of opposing chirality run parallel along a bipolar lateral  junction, and equilibrate their chemical potential during that co"~propagation.
For our Hall bar device, at a given magnetic field, the number of edge channels (filling factors) in the region I and II\&III (II\&III are set to the same gate potential in the following) can be changed independently by the gate voltages $V_{GI}$ and $V_{GII\&III}$.
The number of edge states in each of the regions is $\nu_{I}$ and $\nu_{II} (= \nu_{III})$ (the sign indicates the chirality).
The transmission problem can be solved by the Landauer"~Büttiker formalism \cite{Buttiker.1986,Buttiker.1988, Abanin.2007}.
The four probe resistance $R_{4p} = R_{SD,14}$ in unipolar configuration (same carrier type/chirality in all regions, $nn'n$ and $pp'p$) is calculated to be 
\begin{equation}
    R_{4p,uni} = \lvert \dfrac{1}{|\nu_I|} - \dfrac{1}{|\nu_{II}|} \rvert \cdot \dfrac{h}{e^2},
    \label{unipolar}
\end{equation}
while in bipolar configuration (opposite carrier type/chirality in region I and regions II\&III, $npn$ and $pnp$)
\begin{equation}
    R_{4p,bi} =  \lvert \dfrac{1}{|\nu_I|} + \dfrac{1}{|\nu_{II}|} \rvert \cdot \dfrac{h}{e^2}
    \label{bipolar}
\end{equation}
is expected.

The edge channel transmission measurement for our device is shown in Fig.\,\ref{Fig:3}, plotting the four probe resistance as a color"~map against the two gate voltages $V_{GI}$ and $V_{GII\&III}$.
Fig.\,\ref{Fig:3}a, measured at 6.5\,T, shows quantized edge channel transmission in four quadrants: $nn'n$, $npn$, $pp'p$ and $pnp$.
The quadrants are indicated in Fig.\,\ref{Fig:3}a and refer to the carrier types of the three separately gated regions of the Hall bar ({cf.} insets of the graph).
All four combinations of $\nu_{I} = \pm 1$ and of $\nu_{II} = \pm 1$ reveal quantized resistance, which is $R_{4p} = 0$ in the unipolar regime ($nn'n$ and $pp'p$) and $R_{4p} = 2 h/e^2$ in the bipolar regime  ($npn$ and $pnp$).
The observed resistance values precisely match  Eqs.\,(\ref{unipolar}) and (\ref{bipolar}). 
While  resistance quantization in the $nn'n$ and $npn$ quadrants has been observed in Refs.\,\cite{Gusev.2013, Calvo.2017} ($npn$ not being particularly well quantized in either of the two), quantization in $pp'p$ and $pnp$ was not observed previously in mercury telluride devices.

The selective transmission of multiple edge channels of the same chirality ($n$ type) are shown in the first quadrant of Fig.\,\ref{Fig:3}a, revealing additional quantized plateaus in the $nn'n$ regime.
A richer variety of plateaus is observed in Fig.\,\ref{Fig:3}b, as the reduced magnetic field (3.0\,T) yields a larger number of filling factor combinations in the $nn'n$ quadrant.
Forty quantized plateaus are observed, forming a checkerboard pattern.
Every plateau corresponds to a fraction of $h/e^2$, as indicated in the line cuts.
Again, Eq.\,(\ref{unipolar}) precisely predicts the quantized resistance values.
The latter is illustrated by the subplot in the upper right corner, that shows a calculation of the resistance map as a function of the carrier density in the gated regions.
The edge channel number is selected to increase at half filling factor, which fits the experimental observations best.
The checkerboard pattern including its amplitude is precisely matched by the calculation.
To compare, in Ref.\,\cite{Calvo.2017} the quantization is already significantly altered when two edge channels are used in the back gated regions (equivalent to region II\&III here). 

Overall, the observed quantum Hall edge channel transmission fits theory significantly better over an extended parameter space than in previous works.
We attribute this to the sharp density gradients created by the overlapping gates.
They make it possible to make devices that do not need a back gate, and thereby allow for lattice"~matched substrates  that induce significantly less disorder.
Simultaneously,  compromises regarding edge channel separation  due to an ungated region between neighboring top gates (which may affect edge state transmission/equilibration) are avoided.

\section{Final Remarks}

In this work we discuss overlapping top gate electrodes that allow for gate defined lateral junctions with a minimized separation between neighboring gates, employing low temperature atomic layer deposited hafnium oxide as a gate dielectric.
A device based on a mercury telluride quantum well reveals linear and decoupled gate action under the neighboring gates along with a dielectric breakdown field of order $>1 \cdot 10^9$\,V$/$m.
The design proves to be robust to gate shorts in fabrication.
Poisson calculations show the significant advantage of overlapping gates compared to conventional split gates to avoid ungated areas in lateral junctions.
We observe quantized selective quantum Hall edge channel transmission in all four quadrants of the junctions $nn'n$, $npn$, $pp'p$ and $pnp$, the latter two of which have not been observed in mercury telluride before.
In general, the observations precisely match the established theory, and are significant improvements over prior works on this phenomena in the material system.
We attribute the improvements to the overlapping gates. 

The overlapping gate technique is applicable to any material class such as III-V and II-VI semiconductors, graphene, or 2D materials, in particular for (but not limited to)  the formation of gate defined lateral junctions.
As the low"~temperature atomic layer deposition process employed here supports a very gentle heat load, it is of particular use for heat sensitive materials and materials where back gating is not available.
The deposition of the first dielectric could be carried out \textit{in situ} after material growth, so the approach is compatible with air sensitive materials.
Other dielectrics, such as aluminum oxide or zirconium oxide, are available for the process, further widening its range of applications.
The process is fully compatible with optical lithography (mask"~based or mask"~less), even at smallest gate separations, allowing for fast and efficient prototyping and process integration as electron beam lithography can be fully avoided.

\section*{Acknowledgements}
We acknowledge financial support from the Deutsche Forschungsgemeinschaft (DFG, German Research Foundation) in the project SFB 1170 (Project ID 258499086) and in the Würzburg-Dresden Cluster of Excellence on Complexity and Topology in Quantum Matter \textit{ct.qmat} (EXC 2147, Project ID 390858490), and from the Free State of Bavaria (Institute for Topological Insulators).

\bibliography{Literature.bib}
\end{document}